\documentclass[preprint,amsmath,amssymb,superscriptaddress,aps,pra,a4paper,floatfix]{revtex4-2}
\usepackage{graphicx}
\usepackage[paperwidth=212mm,paperheight=297mm,centering,hmargin=1.65cm,vmargin=2.6cm]{geometry}
\usepackage{color}
\usepackage{dcolumn}% Align table columns on decimal point
\usepackage{bm}% bold math
\usepackage{physics}
\usepackage{float}

\usepackage{chemformula}
\usepackage{natbib}

\begin{document}

\title{Bose-Einstein Condensation of Light in a Semiconductor Quantum Well Microcavity}

\author{Ross~C.~Schofield}
\affiliation{Blackett Laboratory, Imperial College London, Prince Consort Road, SW7 2AZ, London, United Kingdom}

\author{Ming~Fu}
\affiliation{Blackett Laboratory, Imperial College London, Prince Consort Road, SW7 2AZ, London, United Kingdom}

\author{Edmund Clarke}
\affiliation{EPSRC National Centre for III-V Technologies, University of Sheffield, S1 3JD, UK }

\author{Ian Farrer}
\affiliation{EPSRC National Centre for III-V Technologies, University of Sheffield, S1 3JD, UK }

\author{Aristotelis Trapalis}
\affiliation{EPSRC National Centre for III-V Technologies, University of Sheffield, S1 3JD, UK }

\author{Himadri~S.~Dhar}

\affiliation{Department of Physics, Indian Institute of Technology Bombay, Powai, Mumbai 400076, India}

\author{Rick~Mukherjee}
\affiliation{Blackett Laboratory, Imperial College London, Prince Consort Road, SW7 2AZ, London, United Kingdom}
\affiliation{Center for Optical Quantum Technologies, Department of Physics, University of Hamburg, Luruper Chaussee 149, 22761 Hamburg, Germany}

\author{Jon Heffernan}
\affiliation{EPSRC National Centre for III-V Technologies, University of Sheffield, S1 3JD, UK }

\author{Florian~Mintert}
\affiliation{Blackett Laboratory, Imperial College London, Prince Consort Road, SW7 2AZ, London, United Kingdom}
\affiliation{Helmholtz-Zentrum Dresden-Rossendorf, Bautzner Landstraße 400, 01328 Dresden, Germany}

\author{Robert~A.~Nyman}
\affiliation{Blackett Laboratory, Imperial College London, Prince Consort Road, SW7 2AZ, London, United Kingdom}
\affiliation{Oxford Ionics Limited, Begbroke Science Park, Begbroke OX5 1PF, United Kingdom}

\author{Rupert~F.~Oulton}
\affiliation{Blackett Laboratory, Imperial College London, Prince Consort Road, SW7 2AZ, London, United Kingdom}

\date{\today}
\begin{abstract}
{\bf When particles with integer spin accumulate at low temperature and high density they undergo Bose-Einstein condensation (BEC). Atoms \cite{Davis1995,Anderson1995}, solid-state excitons \cite{High2012} and excitons coupled to light \cite{Deng2002} all exhibit BEC, which results in high coherence due to massive occupation of the respective system's ground state. Surprisingly, photons were shown to exhibit BEC much more recently in organic dye-filled optical microcavities \cite{Klaers2010Condensation}, which, owing to the photon's low mass, occurs at room temperature. Here we demonstrate that photons within an inorganic semiconductor microcavity also thermalise and undergo BEC. Although semiconductor lasers are understood to operate out of thermal equilibrium \cite{Siegman1986}, we identify a region of good thermalisation in our system where we can clearly distinguish laser action from BEC. Based on well-developed technology, semiconductor microcavities are a robust system for exploring the physics and applications of quantum statistical photon condensates. Notably, photon BEC is an alternative to exciton-based BECs, which dissociate under high excitation and often require cryogenic operating conditions. In practical terms, photon BECs offer their critical behaviour at lower thresholds than lasers \cite{Bloch2022}. Our study shows two further advantages of photon BEC in semiconductor materials: the lack of dark electronic states allows these BECs to be sustained continuously; and semiconductor quantum wells offer strong photon-photon scattering. We measure an unoptimised interaction parameter, $\mathbf{\Tilde{g}=0.0023\pm0.0003}$, which is large enough to access the rich physics of interactions within BECs, such as superfluid light \cite{Marelic2016} or vortex formation \cite{Dhar2021}.}

\end{abstract}

\maketitle

\section{Introduction}
\noindent  

Semiconductor lasers, first demonstrated in 1962 \cite{Hall1962, Nathan1962}, are now an essential underpinning technology, ubiquitous in research and industry due to their ability to generate bright, coherent and directional radiation from electricity. A conventional understanding of the laser process in a semiconductor involves a population inversion of electrons, excited from the valence to the conduction band, which amplifies its own light emission within an optical resonator. The apparently extreme conditions necessary for optical gain suggest that neither the semiconductor's electrons nor resonant photons can be in thermal equilibrium, either with themselves or each other. Despite this intuition however, steady state optical gain in semiconductors arises from electrons and holes in thermal equilibrium \cite{Bernard1961}. Furthermore, it is well known that a detailed balance of absorption and emission allows electromagnetic radiation to be in thermal equilibrium with its surroundings. Under ambient conditions, Planck's spectrum peaks in the mid-infrared but visible and near infrared light also achieves thermal equilibrium when interacting with suitable light emission materials \cite{Roosbroeck1954,Wurfel1982}. This has been demonstrated for organic dye embedded in and tuned to an optical cavity resonator with a well-defined ground state \cite{Klaers2010}. In this system, with increasing excitation, the photon population reaches a critical number where the chemical potential of light approaches the cavity's ground state energy and Bose Einstein Condensation (BEC) occurs \cite{Klaers2010Condensation}. While offering the high temporal and spatial coherence characteristics of lasers, condensates of light are distinct as they operate robustly in their ground state, exhibit nonlinear and many body physics described by quantum statistical mechanics and do not require carrier inversion, so manifest critical behaviour below laser threshold \cite{Bloch2022,Deng2002}. 

In this work we identify the operating regime of inorganic semiconductor microcavities where their electronic and photonic populations are in thermal equilibrium and thus produce condensates of light at room temperature. Excited using a continuous wave laser, these condensates can be sustained indefinitely, an achievement only recently realised for matter-based condensates \cite{Chen2022}. We study the photon condensate as a function of the light matter coupling strength, allowing us to identify the phases of lasing and condensation, which closely follow theoretical predictions \cite{Hesten2018}. This establishes the underpinning physics of photon condensation in inorganic semiconductor microcavities, confirming evidence of this interpretation in related vertical cavity surface emitting lasers \cite{Barland2021}, and other semiconductor systems \cite{Bajoni2007}.

Our approach addresses key limitations of other optical condensates. Inorganic semiconductor materials have much lower transition rates from bright to long-lived dark states than those of organic dyes, which avoids the requirement for low repetition rate pulsed excitation to avoid the shelving of carriers \cite{Klaers2010Condensation}. Distinct from the condensation of exciton-cavity polaritons \cite{Deng2002,Kasprzak2006,Balili2007,Bajoni2007}, condensates of light are only weakly coupled to their surroundings and so bypass the low temperatures and limited excitation conditions often necessary to sustain bound excitons. Even for materials with stable excitons at room temperature, condensed polaritons dissociate under strong excitation conditions, where they revert to normal laser operation \cite{Deng2003,Bajoni2007,Pieczarka2022}. This enables the use of technically relevant III-V materials despite their low exciton binding energies, as a robust solid state system for exploring the physics and application of quantum statistical condensates.

\section{System characterisation}
Figure 1(a) illustrates the inorganic semiconductor microcavity. One half is a GaAs/AlAs heterostructure DBR mirror and InGaAs quantum well, while the other half is a commercially manufactured DBR mirror on a concave glass substrate with a radius of curvature, $\rho=0.2$ m. The cavity length $\bar{L}$, was locked in position with interferometric stabilisation (see Methods). All measurements were performed with a longitudinal mode number of $q=9$ to ensure the quantum well emits into a single longitudinal mode. The commercial mirror has a reflectivity $R_b>99.995$ \%, while the semiconductor mirror has a reflectivity $R_f>99.95$ \%, so that light is mainly emitted through the GaAs substrate. The cavity loss rate, $\kappa_{q=9} = 1.6 \pm 0.2 \times 10^{10}$~s$^{-1}$ was estimated from laser threshold measurements (see Methods), and is consistent with the mirror reflectivities, which give a loss rate $\kappa > 10^{10}$ s$^{-1}$.

The microcavity defines a 2D photon mode spectrum, which is a function of axial position, $r$, and emission direction ${\bf k}=\{k_r,k_z\}$ \cite{Klaers2010},
\begin{eqnarray} E_{ph}(k_r,r) = mc^2 + \frac{\hbar^2 k_r^2}{2m} + \frac{1}{2}m\Omega^2 r^2 - m c^2 \frac{n_2}{n^3} I(r), \end{eqnarray}\label{eqnTH-1}
where $m=h/c\lambda_{co}$ is the photon mass, $\lambda_{co}$ is the ground state cavity cut-off wavelength, $k_r$ is the transverse wavevector, $\Omega = (c/n)/\sqrt{\bar{L}\rho}$ is the transverse cavity trapping frequency, $n$ is the average cavity refractive index, and $n_2$ is the nonlinear refractive index. The final term accounts for photon-photon interactions due to non-linear refraction, where $I(r)$ is the intra-cavity spatial intensity distribution.

\begin{figure}
    \centering
    \includegraphics[width=0.75\columnwidth]{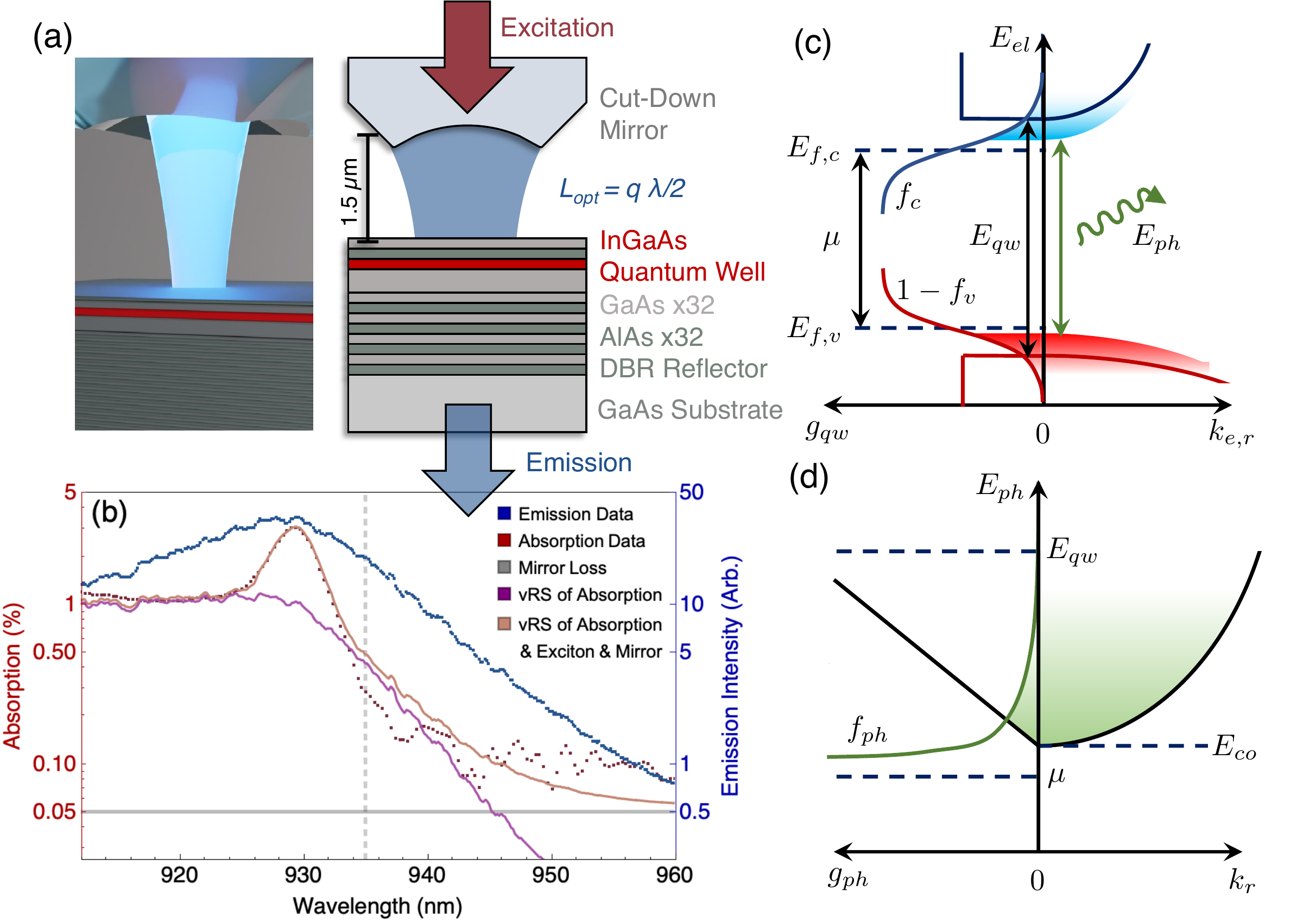}
    \caption{(a) Illustration and schematic of the inorganic semiconductor quantum well microcavity. (b) Quantum well emission spectrum (blue points) and absorption spectrum (red points). The van Roosbroeck-Shockley (vRS) relation transforms the photoluminescence spectrum into the expected absorption spectrum (purple line) for a semiconductor in thermal equilibrium. Correspondence between expected and measured absorption spectra (orange line) is seen when accounting for mirror loss (grey line) and bound exciton absorption at $930$~nm. We expect thermalisation to occur to the right of the dashed vertical line, where absorption and emission follow the vRS relation. (c) Electronic dispersion and thermal population densities following Fermi-Dirac statistics. (d) Photon dispersion and thermal population density following Bose-Einstein statistics.}
    \label{fig:theory}
\end{figure}

The absorption of inorganic semiconductors is controlled through the use of heterostructures, \textit{i.e.} regions of the crystal where the bandgap is varied via composition. Here, we use a single InGaAs quantum well (see Methods) with absorption peak wavelength $\lambda_{qw} = 930$ nm, which is far from the band-egde absorption of the GaAs substrate at $875$ nm. The quantum well sample without commercial mirror was initially characterised by optical excitation with a continuous $785$ nm wavelength laser to verify its suitability for photon condensation. The measured emission and absorption spectra are shown in Fig.~\ref{fig:theory}(b). Intraband carrier-phonon scattering on $10^{-13}$ s timescales ensures that carriers maintain a thermal distribution and so the absorption exhibits a characteristic Urbach tail, emulated also by a long wavelength photoluminescence spectral tail. 

It is well-known that the ratio of absorption and emission in semiconductors at thermal equilibrium follows the van Roosbroeck-Shockley (vRS) relation \cite{Roosbroeck1954,Bhattacharya2012}, analogous to the Kennard-Stepanov relation for organic materials \cite{Kennard1918}. By determining where the vRS relationship holds between the measured emission and absorption, shown in Fig.~\ref{fig:theory}(b), we can identify the spectral regions where thermalisation occurs \cite{Kirton2013}. To do this we apply the vRS relation to the measured photoluminescence to find a predicted vRS absorption spectrum. As the quantum well cannot be isolated from the DBR mirror beneath, the vRS converted absorption and measured absorption diverge for wavelengths greater than $945$~nm. Including the DBR mirror loss restores the correspondence and reveals that good thermalisation of electrons and phonons occurs for $\lambda>935$ nm. At shorter wavelengths another divergence of vRS and measured absorption is attributed to bound exciton absorption, giving the peak visible at $930$~nm \cite{Chemla1984,Yoshita2012,Bhattacharya2015}. At room temperature, excitons in III-V semiconductors are weakly bound and dissociate into free electrons and holes on picosecond timescales \cite{Chemla1984}, compared to the typical nanosecond radiative lifetime of a quantum well. As such they do not meaningfully contribute to the emission spectrum of the quantum well, which is predominantly from the electronic continuum. A Gaussian function models the exciton absorption and reveals the correspondence between vRS and measured absorption across the whole spectral range, shown in Fig.1(b). Thus, we confirm that for $\lambda>935$ nm, carriers and phonons in the quantum well are in thermal equilibrium, a pre-requisite for photon condensation \cite{Kirton2013}. Thermal equilibrium breaks down near the exciton absorption in this III-V system, but could be sustained either under cryogenic conditions or for materials with stable excitons at room temperature.

Semiconductor lasers often use multiple quantum wells to increase gain and power. Here, only a single quantum well produces the necessary detailed balance of cavity absorption, $\alpha$, and mirror loss $\kappa$ to achieve photon thermlisation in the range of photoluminescence \cite{Kirton2015}. Fig.~\ref{fig:theory}(b) shows that the single quantum well provides a peak of $\sim 3$ \% round-trip loss \cite{Fang2013}. Meanwhile, the round trip mirror loss is $1-R_fR_b<6\times 10^{-4}$, indicated by the horizontal line on Fig.~\ref{fig:theory}(b). We define a thermalisation parameter, $\gamma=\alpha/\kappa$ \cite{Hesten2018,Rodrigues2021}, with photon condensation expected in the range $0.1<\gamma<10$. This occurs in the long wavelength Urbach tail of the quantum well absorption between $935$ nm $ <\lambda_{co}< 960$ nm. We have thus designed the semiconductor mirror's central wavelength, $\lambda_{dbr}=950$ nm to overlap with this region. We note that increasing the number of quantum wells would shift the range of good thermalisation to regions where photoluminescence is weaker and would thus require adjusting $\lambda_{dbr}$. This is a departure from the design approach for conventional semiconductor lasers.

Consider now the thermalisation and condensation mechanisms for a semiconductor quantum well in a microcavity. Fig.~\ref{fig:theory}(c) shows the electronic energy spectrum, $E_{el}$ as a function of in-plane electron momentum, $k_{e,r}$, (right) and density of electronic states, $g_{qw}$, (left) under optical excitation. Electron and hole populations are governed by Fermi-Dirac occupancy factors, $f_c$, and $1-f_v$, respectively (see Methods). Their positions are set by quasi-Fermi levels for electrons, $E_{f,c}(N)$, and holes, $E_{f,v}(N)$, and controlled by the density of optically excited carriers, $N$. The net rate of photon production at energy $E_{ph}$ is set by a detailed balance of mode loss, $\kappa s(E_{ph},N)$, absorption, $\alpha (f_c-f_v)s(E_{ph},N)$, and spontaneous emission $A f_c(1-f_v)$, where $s(E_{ph},N)$ is the photon number and $A$ is the spontaneous emission rate. With good photon thermalisation, $\gamma=\alpha/\kappa > 1$, the photon emission spectrum follows a Bose-Einstein distribution (see Methods),

\begin{eqnarray} s(E_{ph},N) \approx g_{ph}\frac{A/\alpha}{e^{(E_{ph}-\mu(N))/k_BT}+1}, \end{eqnarray}\label{eqn-BEC}

where the photon chemical potential is the difference between the electronic quasi-Fermi levels, $\mu(N)=E_{f,c}(N)-E_{f,v}(N)$. Fig~\ref{fig:theory}(d) shows the photonic energy spectrum as a function of $k_r$ (right) and the linear 2D photon density of states $g_{ph}$, with occupancy now set by the Bose-Einstein factor, $f_{ph}$. Photon thermalisation thus follows naturally from the relationship between the Fermi-Dirac and Bose-Einstein occupancy factors, $f_c(1-f_v)=f_{ph}(f_c-f_v)$. As shown in Fig ~\ref{fig:theory}(b), photon thermalisation is expected to occur deep into the semiconductor's Urbach tail, below the quantum well band edge, where $E_{ph}<E_{qw}=hc/\lambda_{qw}$. For increasing carrier density, we would expect the chemical potential to increase, until it reaches the cavity ground state energy, $\mu\simeq E_{co}=hc/\lambda_{co}$, where photon condensation would be expected. We note that this is below the condition for electronic inversion where $\mu \geq E_{qw}$.

\section{Experimental}

\begin{figure}
    \centering
    \includegraphics[width=0.85\columnwidth]{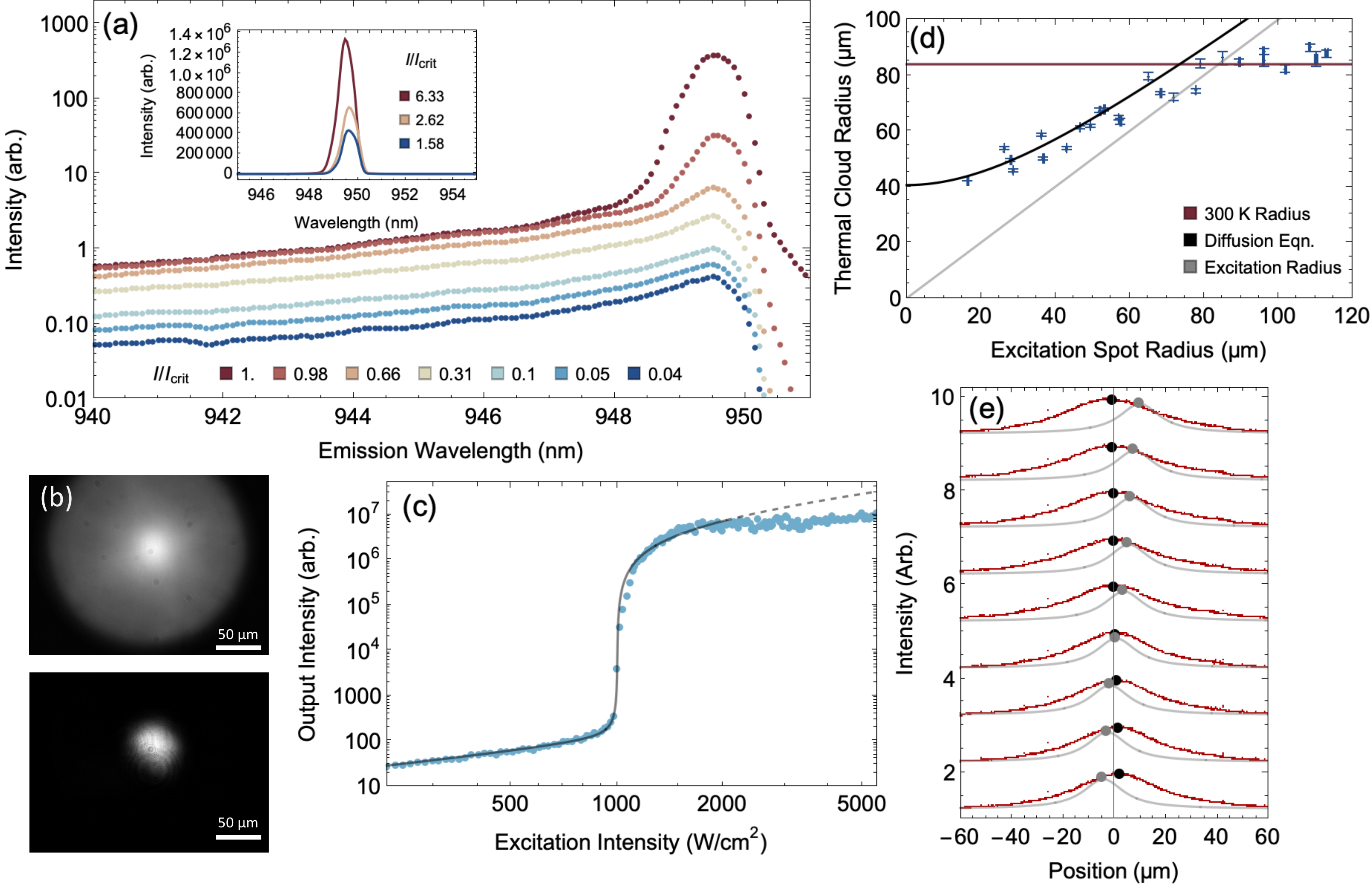}
    \caption{Thermalisation and condensation of light in a semiconductor quantum well microcavity. (a) Emission near the critical excitation intensity,  $I_{crit}$ for a cavity tuned to $\lambda_{co}\approx 950$~nm. Inset shows spectra above $I_{crit}$ in linear scale. (b) Camera images of the thermal cloud for $I < I_{crit}$ (top) and the condensate for $I > I_{crit}$ (bottom). (c) Condensate intensity as a function of excitation intensity. Condensate intensity is found through integrating the data in (a) around $\lambda_{co} \pm 1$~nm. Data is fitted to a semiconductor laser model (see Supplementary Information). Deviation from linear behaviours at high excitation intensity is consisting with heating of the sample. (d) Plot showing the relationship between excitation beam radius and thermal cloud radius. Data is fit with a diffusion relationship $\sqrt{\sigma_p^2 + L_{ph}^2}$ (black), see text. Dashed gray line indicates where excitation and thermal cloud radii are equal. The solid gray line is the expected cloud radius, $84$ \textmu m at room temperature. Thermal cloud size is limited by the diffusion length of $L_{ph}=41\pm3$ \textmu m in the semiconductor and increases with pump spot size until it reaches the expected thermal cloud radius. (e) Plot showing the spatial intensity of the thermal cloud (red) as excitation beam position (gray) is varied. }
    \label{fig:therm}
\end{figure}

Figure~\ref{fig:therm}(a) shows emission from the microcavity where the cut-off wavelength for longitudinal mode order $q=9$ is set to, $\lambda_{co}=950$ nm, corresponding to $E_{co} \approx E_{qw}-1.5 k_BT$. At low pump intensity, we observe a Maxwell-Boltzmann distribution, which is indicative of thermalised photons according to Eqn.\ref{eqn-BEC} \cite{Klaers2010Condensation}. As the pump intensity increases, emission near $\lambda_{co}$ grows faster than the thermal tail, until we reach the threshold intensity, $I_{pc}$. At this point the ground state population becomes dominant, confirmed also by the light's spatial distribution in Fig.~\ref{fig:therm}(b). We observe saturation of the thermal tail with increasing ground state population.%, indicating a large condensate fraction. 
The inset of Fig.~\ref{fig:therm}(a) shows the increase in the population of the lowest energy state with intensities above $I_{crit}$. The condensation is sustained continuously and remains stable under a range of cut-off wavelength and pump conditions.

In order to accurately determine the critical intensity, measured spectra are integrated around $\lambda_{co}$ and plotted versus pump intensity $I$, as shown in Fig.~\ref{fig:therm}(b). A critical intensity of $I_{crit}=1$ kWcm$^{-2}$ was determined. We also measured a maximum continuous output power from the condensate of $\approx 320$ \textmu W, for an input power of $50$~mW. 

\subsection{Thermal Cloud Size}

It is common to verify the photon temperature by fitting the thermal tail in Fig 2a to a Maxwell-Boltzmann factor \cite{Klaers2010}. This approach does not work well for this system, as we find effective temperatures of $T_{eff}\approx 100$ K, depending on spectrometer alignment and slit width. We attribute this low value to the difficulty in measuring higher energy photons at large angles within our system. Looking at Eqn.~\ref{eqnTH-1}, we see that to measure higher energy photons we must collect a wide range of emission angles ($k_r \propto \sin \theta$, for surface normal angle, $\theta$) or use a large field of view ($r$). Due to the high refractive index of GaAs, $n = 3.5$, we collect emission through a low solid angle ($\theta<17^\circ$) set by total internal reflection at the substrate-air interface, further limited by the collection optics' numerical aperture. Similarly, the narrow spectrometer entrance slit limits the field of view of the thermal cloud. These limitations to extracting an effective temperature from spectral measurements were also noted in previous works \cite{Greveling2018,Walker2018}.

Instead, we have used the spatial distribution of photons on a camera to probe the temperature of the thermal cloud. Figure~\ref{fig:therm}(c) shows the observed thermal cloud radius as a function of the excitation beam radius, $\sigma_p$. For $\sigma_p>\sigma_{th}=\sqrt{k_B T q \lambda_0^2 R / 2 h c n}=84\pm2$ \textmu m, the thermal cloud reaches the expected size for the ambient temperature \cite{Marelic2015,Keeling2016}. For our cavity geometry at $q=9$, $n=2.5\pm0.1. $ %SEE WHAT WE'VE ALREADY DEFINED. Here, the cavity effective index value of $n_{eff}\sim 2.5$, is consistent with mirror spacing and effective penetration depth into the DBR mirror \cite{Koks2021}. 
This confirms photon thermalisation in the microcavity at room temperature below the critical pump intensity. For $\sigma_p<\sigma_{th}$, the thermal cloud size follows a diffusion rule, $\sigma_{th} = \sqrt{\sigma_p^2 + L_{ph}^2}$, which indicates an in-plane photon diffusion length, $L_{ph} = 41\pm3$ $\mu$m. This is consistent with the photon diffusion lengths observed in dye filled microcavities \cite{Nyman2014}. For the rest of this letter, we operate the system at an excitation spot diameter of $20$~\textmu m to minimise the laser power required.

We have also varied the position of the excitation beam relative to the optical axis, as shown in Fig~\ref{fig:therm}(e). The thermal cloud does not shift in position from the cavity's energetic minimum, again confirming thermalisation \cite{Klaers2010}.

\subsection{Condensed phase}
\begin{figure}
    \centering
    \includegraphics[width=0.5\columnwidth]{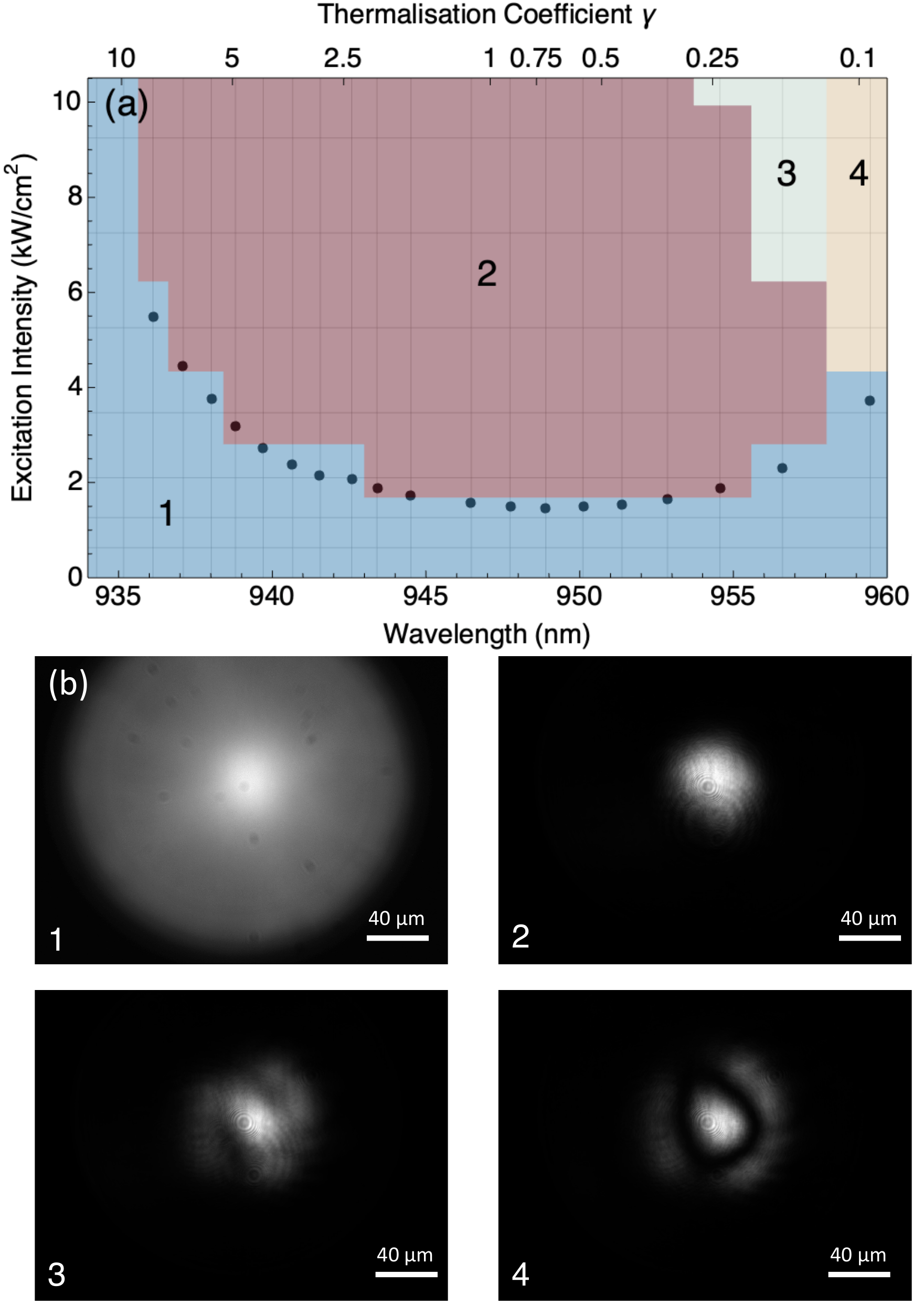}
    \caption{(a) Phase diagram characterising the spatial emission distribution as a function of cavity cut-off wavelength, $\lambda_{co}$, and excitation intensity. Data points indicate the critical excitation intensity, $I_{crit}$, extracted from spectral data. (see Fig.2.~(b) for methodology.) Coloured areas represent the unique regions identified by using a clustering algorithm on images of the emission distributions taken at the values indicated by the intersecting horizontal and vertical lines. The clustering algorithm is given an unsorted list of images and assigns these to groups. Four regions are identified: uncondensed/thermal cloud; photon condensate; break down of single mode condensation; and higher order mode condensation \cite{Hesten2018}. Images are available in the Supplementary Material. (b) Example images from the four regions identified in (a).}
    \label{fig:modes}
\end{figure}

The condensed phase occurs above a critical pump intensity, $I_{crit}(\lambda_{co})$, which depends on cut-off wavelength, as shown by the data in Fig.~\ref{fig:modes}(a). For operation nearer to the peak absorption, $\lambda_{qw}$, a higher critical pump intensity is necessary, as $\mu(N)$ must be increased closer to electronic inversion and the thermalisation parameter, $\gamma \gg 1$. We do not consider condensation below $\lambda_{co}=935$~nm, where excitonic absorption causes thermalisation to break down. For increasing $\lambda_{co}$, cavity loss eventually sufficiently exceeds modal absorption, $\gamma < 0.1$ and so photon thermalisation is expected to break down. It is still possible to observe critical behaviour, but with a higher threshold that suggests a larger value of $\mu(N)$ than would be necessary to observe condensation. This indicates the beginning of the laser regime.

To explore the photon condensation regime in more detail, we studied the spatial distribution of the condensed state as a function of $\lambda_{co}$ and excitation intensity. Photon condensation should exhibit a signature of robust ground state occupation. The phase diagram in Fig.~\ref{fig:modes}(a) categorizes the types of spatial emission pattern, from the examples shown in Fig.~\ref{fig:modes}(b). 

Below the critical pump intensity a thermal cloud is observed for all cut-off wavelengths. Above the critical pump intensity and for $936$~nm$<\lambda_{co} < 957$ nm, condensation in the lowest energy cavity mode is observed. We note the lowest $I_{crit}$ values lie within the condensation region. Meanwhile, for $\lambda_{co}\geq 960$ nm, a higher order mode occurs at the critical intensity, suggesting laser operation. The presence of lasing here, rather than condensation, can be explained by the decrease of $\gamma < 0.1$; there is insufficient absorption to allow the system to thermalise before light is lost from the cavity \cite{Kirton2015}. Of particular interest is the behaviour at $\lambda_{co}=956.5$~nm, where condensation initially occurs, however as the intensity is increased further a multimode distribution is observed. This is consistent with the expected phases of photon BEC from both simulation \cite{Hesten2018} and measurement on dye-based condensates \cite{Rodrigues2021}.

\section{Non-linear effects}
\begin{figure}
    \centering
    \includegraphics[width=0.45\columnwidth]{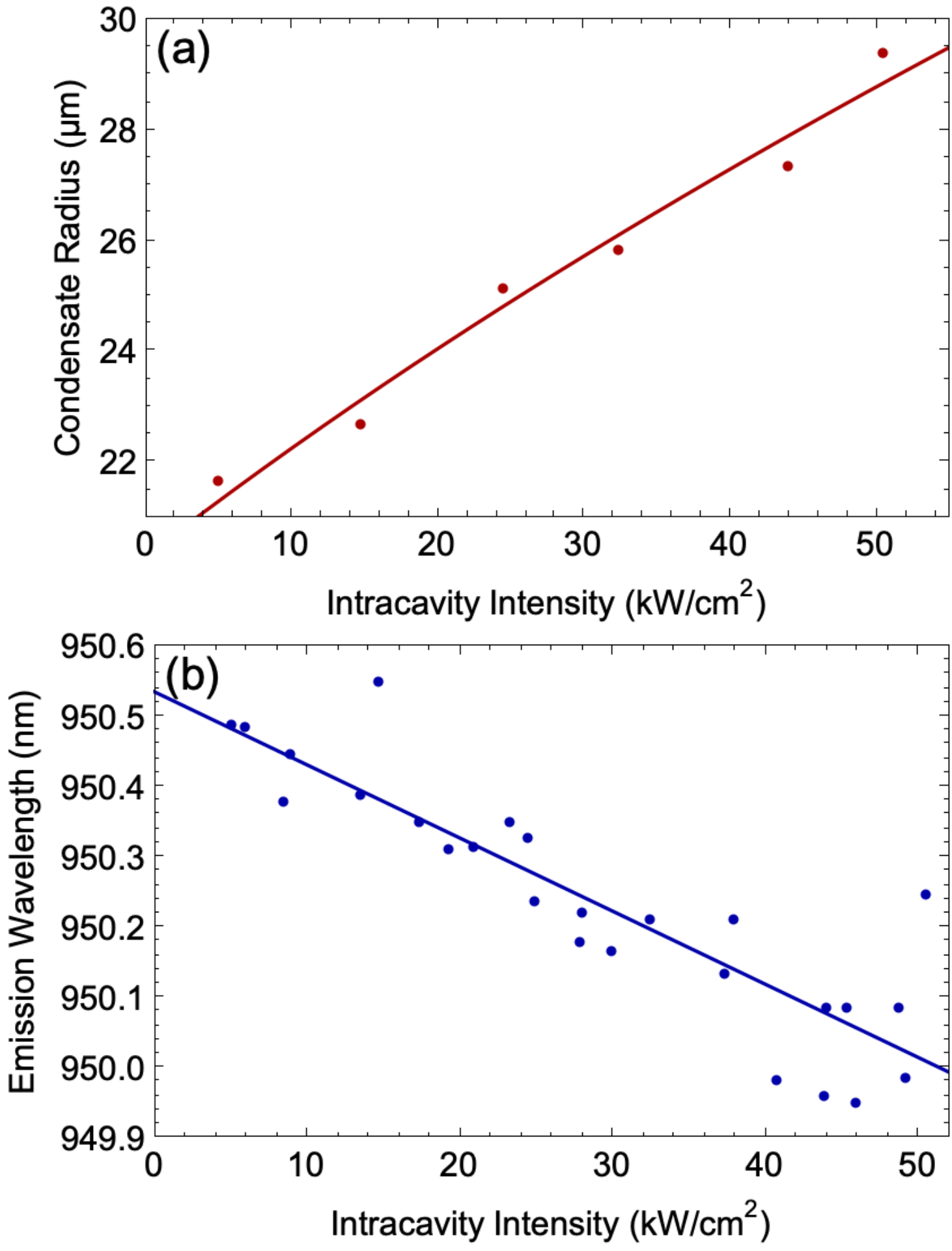}
    \caption{(a) Spatial condensate radius as a function of excitation intensity for $\lambda_{co} = 950$~nm. Repulsive interactions within the condensate, mediated through refractive index modulation, cause the condensate to broaden with increased photon number. Condensate radius $w$ is found from fitting a horizontal slice of the images used in Fig.~\ref{fig:modes}(a) with Gaussian distributions. Data is fit to find $n_2$, from Eqn~IV, giving $\Tilde{g}=0.0024\pm0.0002$. (b) Shift in peak wavelength of condensate emission as a function of excitation intensity for the same $\lambda_{co}$ in (a). The wavelength shift can be seen clearly in the inset of Fig.~\ref{fig:therm}(a). The cavity length is independently locked and stabilised through these measurements. Peak wavelength values come from Gaussian fitting of spectra. The shift to shorter energies is a result of the same refractive index modulation seen in (a). Data is fit to find $n_2$, from Eqn~V, giving $\Tilde{g}=0.0022\pm0.0002$.}
    \label{fig:shift}
\end{figure}

Unlike atom or polariton-based condensates, photon-photon interaction in photon condensates is mediated through weak refractive index variation due to free carrier dispersion, heating or the Kerr effect, leading to relatively small dimensionless interaction parameters, $\Tilde{g}\leq 10^{-4}$ \cite{Klaers2010,Marelic2016}. For dye-based condensates nonlinear refraction is associated with heating of the dye and has an overall repulsive effect where $dn/dT$ is negative; i.e. the refractive index, effective cavity length and cut-off wavelength are all reduced with heating. Evidence of device heating is clearly seen in the saturation of the light output versus pump response in Fig.~\ref{fig:therm}(b). However, $dn/dT$ is positive in GaAs \cite{Talghader1996,Skauli2003} and so heating has an attractive influence on the condensate. However, when looking at the condensate size as a function of pump intensity we see a repulsive effect, as shown in Fig.~\ref{fig:shift}(a). We attribute this refractive index change to the increase in excited carrier density, $N$, where, $dn/dN$, is negative for InGaAs quantum wells \cite{Ehrlich1993,Ehrlich1993InP}. Further evidence for the sign of nonlinear refraction is seen in the central wavelength of the condensate, which shifts to lower wavelengths with intensity, as shown in Fig.~\ref{fig:shift}(b), and also visible within the spectra shown in the inset of Fig.~\ref{fig:therm}(a). We have computed the interaction parameter for the two effects observed (see Methods). Using the condensate size increase we find $\Tilde{g}=0.0024\pm0.0002$, and using the wavelength shift we find $\Tilde{g}=0.0022\pm0.0002$. While these values are higher than those reported for dye based condensates \cite{Klaers2010,Marelic2016}, they are opposed by the effect of heating. The interaction strength could be increased by mitigating heating, either by increasing the excitation beam size or the excitation wavelength. Reducing the longitudinal mode order would also increase the interaction strength, by reducing the air component of the cavity to increase the effective refractive index. It is conceivable that the interaction parameter in the system could approach, $\Tilde{g}\sim10^{-2}$, where superfluid effects \cite{Marelic2016} and vortex creation through stirring \cite{Dhar2021} can be explored.

\section{Conclusion}
In this work we have demonstrated a continuously sustained photon BEC, where thermal equilibrium occurs through light emission and absorption of an inorganic semiconductor quantum well in an open microcavity. The notion that all semiconductor lasers are out-of-equilibrium systems is not necessarily correct. This understanding could enable coherent semiconductor light sources with lower threshold power that lasers, since condensates of light do not rely on electronic inversion. %We found that the quantum well's photoluminescence and absorption follow the van Roosbroek Shockley relation in the Urbach tail region, confirming thermal equilibrium of electrons and phonons here. Photon thermalisation with the quantum well was then seen in both spectral and spatial data at low excitation. With increasing excitation we observed photon BEC, where a large fraction of emitted photons populated the cavity's ground state.
By categorizing photon BEC emission distributions for varying excitation and ground state wavelength, we were able to construct a phase diagram that distinguishes photon BEC from lasing from multimode condensation \cite{Hesten2018,Rodrigues2021}. The range of thermalisation parameters required, $0.1<\gamma<10$, was found to be consistent with the theory of photon BEC developed for dye-based micocavities. We also presented evidence of repulsive non-linear interactions with an unoptimised interaction parameter, $\tilde{g}=0.0023\pm0.0003$, which is already high enough to explore superfluid light \cite{Marelic2016}. Higher than values previously reported for dye-based photon BEC \cite{Klaers2010Condensation,Wurff2014}, we also identified ways to boost this value further towards $\tilde{g}\sim 10^{-2}$. Combining strong photon-photon interactions with ability to sustain a photon condensate continuously brings the exciting possibility to create vortices within the BEC and observe their dynamics \cite{Dhar2021} or look for signs for superfluidity \cite{Marelic2016}.

\section{Methods}

\subsection{Experimental Set-up}
We assembled an open-access microcavity as illustrated in Fig.~1(a). The commercial mirror was cut-down to a diameter of $0.7$ mm, to allow the two halves of the microcavity to approach with a minimum separation of $1.4x$ $\mu$m at the optical axis. We adjusted the cavity length so that the longitudinal mode number, $q = 9$. The optical cavity length, $\bar{L}$, is controlled via a piezo actuator, which is stabilised to the interference pattern of a secondary LED light source. Here, we used the image of interference rings from a reflected $840$ nm laser diode (Thorlabs, LED840L) through the commercial mirror (Layertec), measured on a camera (Blackfly, S BFS-U3-16S2M), to monitor the fluctuating cavity length. This is then used to drive a PID locking loop using a piezoelectric positioning stage (Thorlabs, NFL5DP20M) to lock cavity length to within our spectrometer resolution. 

We excite the quantum well using a 785 nm laser diode (Thorlabs, L785H1) focused through an objective lens (Mitutoyo, M Plan Apo NIR 20X) and the cut-down commercial mirror, although we note we do not fill the back aperture, hence excitation spot sizes larger than the diffraction limit. The emission is collected through the planar mirror using an objective lens (Mitutoyo, M Plan Apo NIR 50X) and split on a beamsplitter between a camera (Blackfly, S BFS-U3-16S2M) and spectrometer (AvaSpec, ULS2048L). We filter the camera image using a 900 nm long pass filter (Thorlabs) to remove any bulk GaAs emission or remaining pump laser.

\subsection{Detailed balance for a semiconductor quantum well in a microcavity}
The net rate of photon production in the $n^{th}$ cavity mode for an optically excited carrier density, $N$, is set by a detailed balance with the modal cavity loss rate, $\kappa_n\approx \kappa$, such that $A_n f_c(1-f_v)+\alpha (f_c-f_v)s_n(E_{ph},N)=\kappa s_n(E_{ph},N)$, where $s_n(E_{ph},N)$ is the photon number, $A_n$ is the modal spontaneous emission rate, and $\alpha_n \approx \alpha$ is the modal absorption rate. $f_{c/v}^{-1}= \exp{(E_{c/v}-E_{f,c/v}(N))/k_BT} + 1$ are Fermi Dirac probability functions, where $E_{c/v}$ are the conduction/valence band extrema energies, and $E_{f,c/v}(N)$ are quasi Fermi levels. With good photon thermalisation, $\gamma=\alpha/\kappa \gg 1$, the photon number in the $n^{th}$ mode follows a Bose-Einstein distribution, $s_n(E_{ph},N) \approx A_n f_{ph}(E_{ph},N)/\alpha_n$, where $f_{ph}(E_{ph},N)^{-1} =  \exp{(E_{ph} - \mu(N))/k_BT} - 1$, with chemical potential, $\mu(N)=E_{f,c}(N)-E_{f,v}(N)$. When combined with the density of photon states, $g_{ph}$, the total photon spectrum follows Eqn.\ref{eqn-BEC} in main text.

\subsection{Intensity curve fitting}
The fluorescence intensity data in Fig.2(a) is fit with
\begin{equation}
    S = \frac{\eta}{2}(1 + \frac{N_{modes} \kappa}{I_{crit}}) I + \frac{\eta}{2}(1 - \frac{N_{modes} \kappa}{I_{crit}})(\sqrt{(I-I_{crit})^2 + 4 I \kappa})- I_{crit}),  
\end{equation}
where $S$ is the integrated count rate, $I$ is the excitation intensity, $I_crit$ is the critical intensity for condensation to occur, $\eta$ is a scaling term corresponding to the collection efficiency, $N_{modes}$ is the number of cavity modes collected, and $\kappa$ is the cavity loss rate. This is a rate equation model accounting for both spontaneous and stimulated emission; the full derivation is provided in the supplementary information. We fit to spectra up to $\approx 2.5$ times saturation intensity as beyond this point excitation beam heating causes deviation from our model, however intensity for all measured spectra are shown for completeness. The output intensity is determined by integrating a $2$~nm slice of spectral data at $\lambda_{co}$.

We find the critical intensity for condensation is relatively low at $I_{c} = 1.01\pm0.01$ kWcm$^{-2}$, which is to be expected for operation below the Bernard-Durrafourg condition. We also find a cavity loss rate $\kappa = 1.6 \pm 0.2 \times 10^{10}$~s$^{-1}$, corresponding to cavity lifetime of $62 \pm 7$~ps. 

\subsection{Image Grouping}
Images of the quantum well fluorescence are grouped using a hierarchical clustering function (FindClusters) in the software Mathematica, with the method set to `Agglomerate'. There is no sorting or labeling of images, with the grouping based entirely on image content. The only constraint on sorting is the number of regions, set to 10, to stop over-sorting.

\subsection{Photon Interaction}
The dimensionless interaction parameter used to quantify interactions within a condensate can be expressed as $\Tilde{g}=-(m_{ph}^4c^6n_2)/(2 \pi\hbar q n)$ \cite{Nyman2014, Klaers2010Condensation}. We can relate the condensate size increase in Fig.~4(a) to the nonlinear refractive index $n_2$ using Eq.~1. We set the energy change from the increase in condensate size, from an initial size of $r_{init}$, according to $1/2 m_{ph} \Omega^2 r^2$ equal to the interaction term $-m_{ph} c^2 n_2/n^3 I(r)$ and rearranging we find
\begin{equation}
    r = r_{init}-\sqrt{\frac{2 c^2 I n_2}{n^3 \Omega^2}}
    %n_2=\frac{1}{2}\frac{\Omega^2 (r_{max}^2-r_{min}^2)n^3}{c^2I(r)}
\end{equation}
which we can fit to measured condensate size. For the change in condensate wavelength we can find the corresponding refractive index change $\Delta n=n_2 I(r)$ through
\begin{equation}
    \lambda = \lambda_{init}+\frac{n_2 \lambda_{init} I}{n}.
\end{equation}
We know the inter cavity intensity $I(r)$ through calibration of spectrometer response with a power measurement of the condensate output, $320\pm40$~\textmu W at highest pump intensity, and output mirror reflectivity.

\bibliography{references}
\end{document}